\documentclass[twocolumn,twocolappendix]{aastex701}
\usepackage{amsmath}

\newcommand{\tno}[1]{TESS~TNO-#1}

\newcommand{\edt}{2026~ED$_{138}$}
\newcommand{\val}[2]{\ifcsname v@#1@#2\endcsname\csname v@#1@#2\endcsname\else\textcolor{red}{??}\fi}
\newcommand{\tval}[1]{\ifcsname t@1@#1\endcsname\csname t@1@#1\endcsname\else\textcolor{red}{??}\fi}
\expandafter\def\csname v@2@H\endcsname{3.6}
\expandafter\def\csname v@2@Dlo\endcsname{520}
\expandafter\def\csname v@2@Dhi\endcsname{1300}
\expandafter\def\csname v@2@Dmid\endcsname{910}
\expandafter\def\csname v@2@r\endcsname{57}
\expandafter\def\csname v@2@galb\endcsname{5}
\expandafter\def\csname v@2@ztfsnr\endcsname{22}
\expandafter\def\csname v@5@arc\endcsname{20}
\expandafter\def\csname v@11@q\endcsname{24.8}
\expandafter\def\csname v@11@Dmid\endcsname{120}
\expandafter\def\csname t@1@aaa\endcsname{30.034}
\expandafter\def\csname t@1@ee\endcsname{0.065}
\expandafter\def\csname t@1@i\endcsname{20.3}
\expandafter\def\csname t@1@nights\endcsname{14}
\expandafter\def\csname t@1@npos\endcsname{28}
\expandafter\def\csname t@1@yfirst\endcsname{2010}
\expandafter\def\csname t@1@ylast\endcsname{2025}
\expandafter\def\csname t@1@ztfsnr\endcsname{32}
\newcommand{\nnew}{eleven}
\newcommand{\Nnew}{Eleven}
\newcommand{\ncand}{eight}
\newcommand{\Ncand}{Eight}
\newcommand{\nlinked}{23}
\newcommand{\nruledout}{three}
\newcommand{\candlocal}{6.6--9.7}
\newcommand{\nsec}{95}
\newcommand{\nccd}{572}
\newcommand{\nknown}{314}
\newcommand{\nknowntwo}{176}
\newcommand{\nknowntno}{294}
\newcommand{\nknowncen}{20}
\newcommand{\Npre}{Nine}
\newcommand{\vhalf}{21.6}
\newcommand{\nappear}{19{,}116}
\newcommand{\nbright}{113}
\newcommand{\nbrightrec}{110}
\newcommand{\rangeV}{20.8--22.1}
\newcommand{\ranger}{25--57}
\newcommand{\rangei}{$16^\circ$--$27^\circ$}
\newcommand{\rangelat}{$9^\circ$--$26^\circ$}
\newcommand{\rangearc}{2--20}
\newcommand{\rangenights}{1--13}
\newcommand{\rangenblind}{two to five}
\newcommand{\npositions}{195}
\newcommand{\yfirstground}{2003}
\newcommand{\ylastground}{2025}
\newcommand{\posmax}{8.9}
\newcommand{\nknowndet}{655}
\newcommand{\nknownpass}{515}
\newcommand{\knownpasspct}{79}
\newcommand{\ncontrolhighi}{three}
\newcommand{\nchancelow}{4.3}
\newcommand{\nlinksmid}{ten}
\newcommand{\nchancemid}{1.2}
\newcommand{\nlinksstrong}{57}
\newcommand{\ndet}{4{,}064}
\newcommand{\ndetcontrol}{3{,}610}
\newcommand{\npairs}{56{,}809}
\newcommand{\nlinks}{67}
\newcommand{\nlinksconf}{38}
\newcommand{\nlinkscand}{10}
\newcommand{\nlinksruled}{four}
\newcommand{\ncontrolpairs}{184{,}550}
\newcommand{\ncontrollinks}{18}
\newcommand{\nchance}{5.5}
\newcommand{\nknownpairs}{847}
\newcommand{\nknownlinks}{669}
\newcommand{\knownlinkpct}{79}
\newcommand{\ntwoknownpairs}{13{,}038}
\newcommand{\ntwoknownlinks}{10}
\newcommand{\nvetoed}{61}
\newcommand{\loonights}{73}
\newcommand{\loomedian}{0.09}
\newcommand{\loowithin}{68}
\newcommand{\ensmax}{0.2}
\expandafter\def\csname v@2@loomax\endcsname{4.0}
\expandafter\def\csname v@2@loolargest\endcsname{1.4}
\expandafter\def\csname v@4@loomax\endcsname{2.8}
\expandafter\def\csname v@5@loomax\endcsname{3.3}
\expandafter\def\csname v@7@chipert\endcsname{2}
\expandafter\def\csname v@7@nnum\endcsname{18}
\expandafter\def\csname v@9@loomax\endcsname{2.5}
\expandafter\def\csname v@9@loolargest\endcsname{1.3}
\expandafter\def\csname v@10@loolargest\endcsname{2.8}
\expandafter\def\csname v@2@loosampled\endcsname{2.9}
\expandafter\def\csname v@2@looacross\endcsname{0.06}
\newcommand{\nztffit}{five}
\newcommand{\ztffitmin}{7.8}
\newcommand{\ztffitmax}{26.5}
\newcommand{\ztfsearchmin}{8.2}
\newcommand{\ztfsearchmax}{27.2}
\newcommand{\ztfoffmax}{0.10}
\newcommand{\troclones}{200}
\newcommand{\troamp}{8.1}
\newcommand{\troamplo}{0.2}
\newcommand{\troamphi}{0.1}
\newcommand{\troperiod}{9.3}

\shorttitle{TNOs discovered by TESS}
\shortauthors{Han et al.}

\begin{document}

\title{\Nnew\ Trans-Neptunian Objects and \Ncand\ Candidates Discovered by TESS}

\author[0000-0002-7127-7643]{Te Han}
\affiliation{MIT Kavli Institute for Astrophysics and Space Research, Massachusetts Institute of Technology, Cambridge, MA, USA}
\email{tehan@mit.edu}

\author[0000-0002-4371-3460]{Jeroen Audenaert}
\affiliation{MIT Kavli Institute for Astrophysics and Space Research, Massachusetts Institute of Technology, Cambridge, MA, USA}
\email{jeroena@mit.edu}

\author[0000-0002-1836-3120]{Avi Shporer}
\affiliation{MIT Kavli Institute for Astrophysics and Space Research, Massachusetts Institute of Technology, Cambridge, MA, USA}
\email{shporer@mit.edu}

\author[0000-0002-5788-9280]{Daniel Muthukrishna}
\affiliation{MIT Kavli Institute for Astrophysics and Space Research, Massachusetts Institute of Technology, Cambridge, MA, USA}
\affiliation{Center for Astrophysics $\vert$ Harvard \& Smithsonian, 60 Garden Street, Cambridge, 02138, MA, USA}
\email{danmuth@mit.edu}

\author[0000-0003-2058-6662]{George~R.~Ricker}
\affiliation{MIT Kavli Institute for Astrophysics and Space Research, Massachusetts Institute of Technology, Cambridge, MA, USA}
\email{grr@mit.edu}

\correspondingauthor{Te Han}
\email{tehan@mit.edu}

\begin{abstract}
We report a blind search for trans-Neptunian objects (TNOs) in Transiting Exoplanet Survey Satellite (TESS) full-frame images from 2018 July 25 through 2026 November 6. We search \nccd\ CCD fields near the ecliptic across \nsec\ TESS sectors. After modeling and subtracting Gaia stars with the TESS-Gaia Light Curve package, we stack the residuals along curved trial paths and link detections across sectors. We identify \nnew\ new TNOs, each detected blindly in \rangenblind\ TESS sectors and independently in archival ground-based images, with arcs of \rangearc~years. They have $V = $ \rangeV, heliocentric distances of \ranger~au, inclinations of \rangei, and diameters of about \val{11}{Dmid}--\val{2}{Dmid}~km for an assumed albedo of 0.08. They lie \rangelat\ from the ecliptic, and the largest is $\val{2}{galb}^\circ$ from the Galactic plane. \Ncand\ additional objects linked across two or three TESS sectors lack secure independent recovery in the archives examined and remain candidates. The search recovers \nknown\ known TNOs and Centaurs, with an empirical recovery fraction of 50\% at $V = \vhalf$ before quality cuts. \Npre\ were discovered after TESS imaged them, including \edt\ 6.8~years before its discovery by Rubin. These results demonstrate the effectiveness of TESS for finding bright outer Solar System objects and extending their observational arcs.
\end{abstract}


\section{Introduction}

\begin{figure*}
\centering
\includegraphics[width=\textwidth]{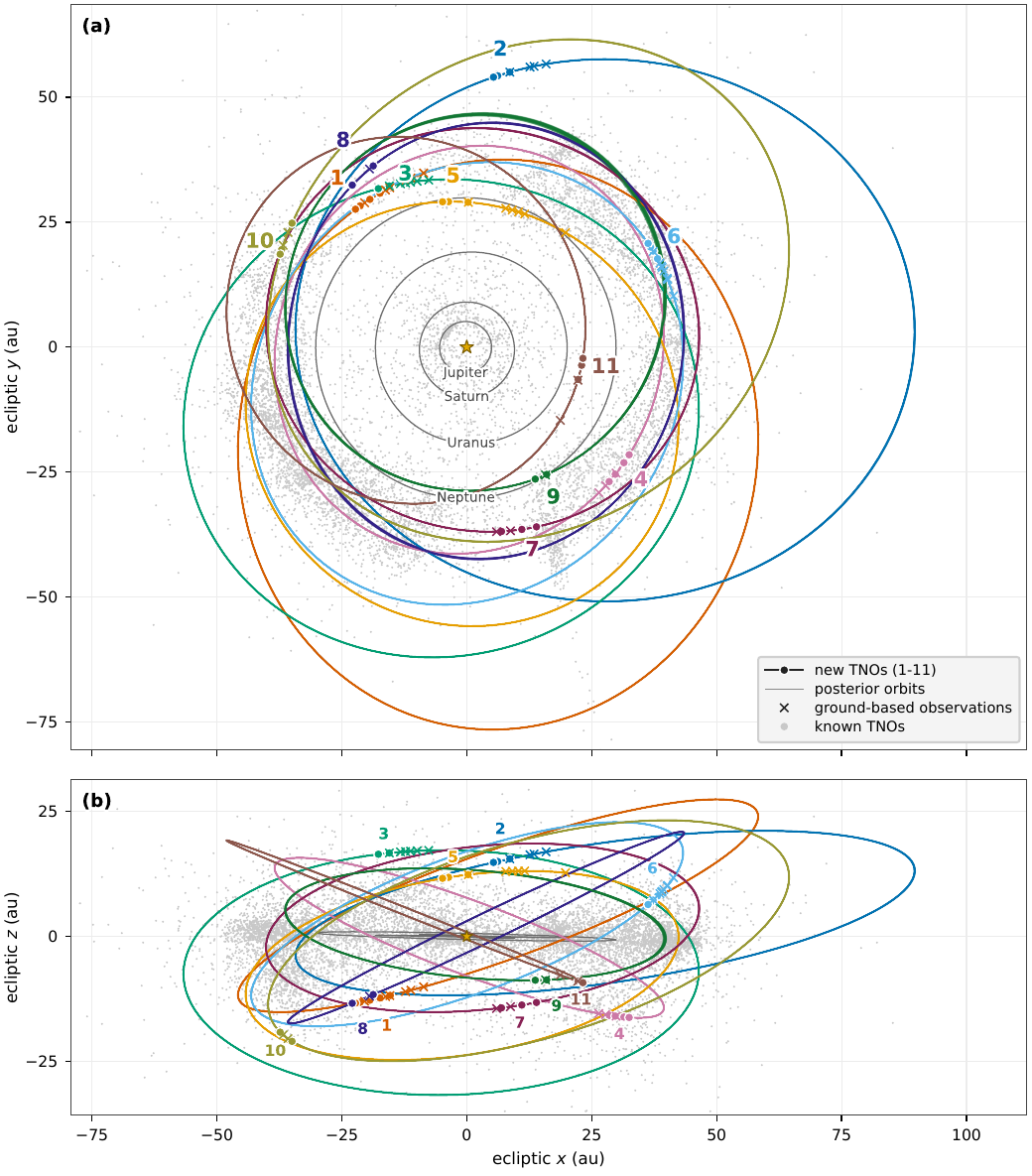}
\caption{Orbits of the \nnew\ new TNOs seen from (a) above and (b) along the ecliptic, on the same scale. Each curve is a fitted orbit with 50 posterior draws, which nearly coincide at this scale, because the orbits are precisely measured. Points mark TESS detections and crosses ground-based observations; numbers follow Table~\ref{tab:new}. Gray points are known TNOs and Centaurs.\label{fig:orbits}}
\end{figure*}

The trans-Neptunian region provides a record of how the giant planets formed and migrated, and its census is still growing. Of the 498 known trans-Neptunian objects (TNOs) with absolute magnitude $H \le 6$, 102 received their designations in 2025 and 2026 (JPL Small-Body Database, 2026 November). The Vera C.\ Rubin Observatory \citep{Ivezic2019} also began to survey the sky and published its first dwarf-planet candidate, \edt\ \citep{MPEC2026U85,Holman2026}.

The Transiting Exoplanet Survey Satellite \citep[TESS;][]{Ricker2015} has four cameras, each with four CCDs that cover $12^\circ \times 12^\circ$, which together image a $24^\circ \times 96^\circ$ strip of sky. It takes a full-frame image (FFI) of the strip every 200--1800~s and stares at it for one sector of about 27~days (about 55~days in Sectors~97 and 98). Since 2018 TESS has covered nearly the whole sky, and it returns to most fields about every two years. A single FFI reaches a depth of only $V \approx 16$--17. A month of FFIs stacked along an object's path reaches $V \approx \vhalf$ (Section~\ref{sec:known}), but the path of an undiscovered object is not known, so a blind search has to try every path that such an object could take. Distant objects are well suited to this search methodology, because they move only a few TESS pixels ($21''$) per day and can remain in one camera for the whole sector.

TESS images have been searched for moving objects before. \citet{Woods2021} detect asteroids in the FFIs without prior ephemerides. For the much fainter TNOs, \citet{Holman2019} and \citet{Payne2019} showed that stacking a sector of FFIs along trial paths can reach about 22nd magnitude and predicted hundreds of detectable objects. \citet{RiceLaughlin2020} searched two sectors along the Galactic plane in this way and listed 17 candidates at 80--200~au, but follow-up of two of the most distant ones with the 4.2~m William Herschel Telescope did not recover them \citep{delaFuenteMarcos2022}. Later pipelines recovered known TNOs \citep{Ganapathi2022} and detected more than a thousand faint asteroids \citep{Nguyen2024} in TESS images. From the ground, stacking six years of images of the Zwicky Transient Facility \citep[ZTF;][]{Bellm2019} along trial orbits recovers known objects without their ephemerides \citep{GeringerSameth2025}. However, no confirmed new TNO has previously come from TESS. Near the Galactic plane a single TESS pixel holds several Gaia stars, whose residuals dominate the stacks. We recently showed that modeling and subtracting Gaia DR3 stars \citep{Gaia2016,Gaia2023} with the TESS-Gaia Light Curve package \citep[TGLC;][]{Han2023} revealed \edt\ in all five TESS sectors that observed it, even at 7--21 Gaia sources per pixel \citep{Han2026}.

In this Letter we extend the precovery of \edt\ to a blind search of the TESS archive for TNOs and discover \nnew\ new TNOs (Figure~\ref{fig:orbits}; Table~\ref{tab:new}), which we call \tno{1} to \tno{11}. Section~\ref{sec:methods} describes the method, Section~\ref{sec:results} presents the TNO survey results, and Section~\ref{sec:discussion} discusses why these objects had been missed and how the TESS archive complements Solar System surveys such as Rubin.

\section{Methods}\label{sec:methods}

We designed the search for objects with $V \lesssim 21.5$ that move 1--6 TESS pixels per day. These are too faint to see in a single FFI but can be detected by stacking a sector of FFIs. We accept prograde orbits with inclinations up to $60^\circ$. These rates correspond approximately to heliocentric distances of 22--150~au if calculated near opposition. TESS's usual footprint at ecliptic latitudes $6^\circ$--$30^\circ$ favors inclined objects, and several TESS sectors pointed along the ecliptic also sample the low-inclination belt.

\subsection{Images, star models and cleaning}\label{sec:data}

We use all available FFIs calibrated by the TESS Image CAlibrator \citep[TICA;][]{Fausnaugh2020} from the start of the mission through 2026 November 6, including a partial Sector~110, and select CCDs with at least one fifth of their area within $30^\circ$ of the ecliptic. In the 18 sectors that we searched first, between 56 and 88, we used all CCDs of cameras 1 and 2, which adds 76 fields farther from the ecliptic. This results in \nccd\ CCD fields in \nsec\ sectors. TGLC models each image as the sum of all Gaia DR3 sources \citep{Gaia2016,Gaia2023}, each at its Gaia position with its predicted TESS flux, convolved with an effective point spread function (ePSF) that is fitted to the image itself, plus a smooth background \citep{Han2023}. Subtracting this model leaves noise, imperfectly modeled stars, remaining scattered light, and also moving objects.

The search works on 3~hr bins, over which the fastest searched object moves 0.75~pixel. In the first 18 sectors we fitted the model to every FFI and averaged the residuals. We then compared this with fitting the model to the FFIs averaged over 3~hr: the residual noise differs by only 0.3\%, because the model is linear in the ePSF, and the number of fits drops by a factor of 6--54, depending on the FFI cadence. We adopted this for all other sectors.

To suppress stationary residuals, we subtract from each pixel its temporal median over the bins 1.5--3 days earlier and later, a ``donut'' template in the sense of \citet{Alexandersen2026}. Spatial high-pass filtering and column-median subtraction remove remaining background structure. We assign inverse-variance weights and discard bins dominated by scattered light. Slow sources lose flux in the median subtraction, but a simulated point source keeps 90\% of its flux at 1~pixel per day and over 99\% above 1.5~pixels per day.

\subsection{Shift and stack}

At each position we search about 11{,}500 velocities: 1--6~pixels per day within $30^\circ$ of the parallactic direction, in steps of 0.04~pixel per day in both components. Each path includes the curvature of a circular orbit with the corresponding mean motion, which can be calculated analytically. For each path we sum the weighted bins under a Gaussian matched filter ($\sigma = 0.8$ pixel) and normalize the S/N by the scatter of random paths in the field. A GPU searches one CCD in one sector in about 90~s (NVIDIA H200).

\subsection{Detections and false alarms}\label{sec:falsealarms}

We keep paths with S/N~$\geq7$: among the $5\times10^{10}$ trial paths of a field, Gaussian noise alone would exceed this level at most 0.06 times. Every such detection must pass automated tests. We split each sector into six segments of about four days and require the detection's six fluxes to agree ($\chi^2$ per degree of freedom below 2.5, allowing 30\% scatter in addition to measurement noise) and the detection to keep 60\% of its S/N without its strongest segment. 98\% of the detections of known objects pass this test. We also reject stationary signals, bright-star residuals and edge artifacts. The measured motion must admit a prograde orbit at 8--400~au with inclination $\leq60^\circ$. We also reject \nvetoed\ detections repeated at the same detector row and motion across multiple CCDs, most likely an instrumental systematic. Finally we renormalize the S/N by the scatter of the paths within 50~pixels (the local S/N, typically 0.84 of the field-wide value; Tables~\ref{tab:detections} and \ref{tab:candidates} list it for the objects reported here).

A pair of detections needs one clear member, which fixes where a weaker partner can be: detections with local S/N~$\geq5.5$ enter the cross-match between TESS sectors (Section~\ref{sec:linking}), and one member of each pair must exceed 6.5. A control search along the time reverse of every path, with the same tests and limits, counts the false alarms. It yields \ndetcontrol\ detections against the \ndet\ of the search that match no known TNO or Centaur, and no S/N limit can make a single-sector detection reliable. For the cross-match the control predicts \nchance\ chance links among the \npairs\ pairs of new detections, \nchancelow\ of them with a member below 6.5, where we find no link (Appendix~\ref{sec:diagnostics}).

Some artifacts pass these tests at high S/N. In the ecliptic sectors, electronic crosstalk copies Uranus and Neptune onto other parts of their CCDs, and main-belt asteroids near their stationary points, which those sectors see $50^\circ$--$60^\circ$ from opposition, move as slowly as TNOs do for a few weeks. Cross-sector orbital consistency and independent ground evidence are therefore both required for confirmation.

\subsection{Cross-matching TESS sectors}\label{sec:linking}

The cross-match is between detections in different TESS sectors; ground-based data are compared with the resulting orbits only afterwards (Section~\ref{sec:confirmation}). Each sector supplies a position at its reference epoch and the mean sky motion of the detected track. For each geometrically compatible pair of sectors we solve Lambert's problem \citep{Bate1971}, varying the two observer-to-object distances. The resulting orbit predicts four motion components, which we compare with the measurements using uncertainties of 0.05~pixel per day (the 0.04~pixel per day step of the velocity grid plus measurement error; pairs of detections of one known object scatter half as much). We accept $\chi^2<4$ for two nominal degrees of freedom, requiring a bound prograde orbit with $e<0.9$, perihelion beyond 15~au and inclination $\leq60^\circ$.

We measure the odds of a chance link in two ways. A pair that includes a control detection cannot be real: \ncontrollinks\ of \ncontrolpairs\ such pairs meet the criterion (Appendix~\ref{sec:diagnostics}), one in 10{,}000, and each of them has a member with local S/N below 7.2. Two different real objects can also link by chance, at any S/N: \ntwoknownlinks\ of \ntwoknownpairs\ pairs of different known objects do, one in 1{,}300. A link therefore remains a candidate until independent archival detections confirm it. We exclude \nruledout\ linked objects because ZTF stacks do not recover the signals expected from their TESS brightnesses (Section~\ref{sec:confirmation}; Appendix~\ref{sec:ztf}). All of their detections have local S/N above 8, so a higher TESS limit alone would not have removed them.

\subsection{Archival confirmation}\label{sec:confirmation}

We searched five public archives along the paths predicted by the posterior of each object's TESS orbit. Two give cataloged single-exposure detections: the NOIRLab Source Catalog DR2 \citep[NSC;][]{Nidever2021}, from Dark Energy Camera \citep[DECam;][]{Flaugher2015}, Mosaic-3 and 90Prime \citep{Dey2019} images of 2012--2019, and the Pan-STARRS1 \citep{Chambers2016} DR2 detections of 2009--2015 \citep{Flewelling2020,Magnier2020}. We kept detections inside the posterior that the catalog does not see repeatedly at one position, as it does stars; Pan-STARRS1 detections must also form pairs within a night that move as the orbit does. In two others we measured positions ourselves: on pipeline-calibrated DECam images from the NOIRLab Astro Data Archive, using their astrometry, and on Canada--France--Hawaii Telescope (CFHT) MegaCam \citep{Boulade2003} images found with the Solar System Object Image Search \citep{Gwyn2012}, tied to Pan-STARRS1 stars in the field. A source must be absent from its position in the images of other nights, which excludes stars and galaxies. In the fifth archive, ZTF, the objects are below the depth of single exposures, so we stack its public difference images \citep{Masci2019} along the predicted tracks, as in the TESS images (Appendix~\ref{sec:ztf}). A stack gives a detection without single positions, so it enters the orbit fits as images, like TESS (Section~\ref{sec:orbits}); for \tno{2}, \tno{3}, \tno{5} and \tno{6} the stacks also narrowed the orbits with which we searched the other archives.

We then fit one orbit to all candidate positions, with the TESS posterior as a prior, by least squares that down-weight outliers, drop positions more than both $4\sigma$ and $0.5''$ from the orbit, and refit until the set no longer changes. Table~\ref{tab:ground} lists the positions that remain, on \rangenights\ nights for each of the \nnew\ TNOs.

\subsection{Orbits}\label{sec:orbits}

The orbit calculations use Barycentric Dynamical Time (TDB) epochs, taking TICA exposure midpoints directly and converting Pan-STARRS1 TAI and other ground-based UTC midpoints to TDB.

Each linked object first gets a two-body orbit from TESS alone, which only supplies the tracks for the archival searches. For \tno{1} to \tno{8} we fitted it directly to the TESS images; for \tno{9} to \tno{11} and the candidates we drew it from the orbits that the TESS cross-match allows, according to the $\chi^2$ of the linked pairs. In the image fit, for nonnegative independent sector fluxes and fixed Gaussian noise, the matched-filter profile likelihood is $\ln L=\tfrac12\sum_s\max(\mathrm{S/N}_s,0)^2$. We parametrize the orbit by the object's position and distance in two of its sectors, include the light travel time \citep{BernsteinKhushalani2000}, and sample the posterior with \texttt{emcee} \citep{ForemanMackey2013}.

All \nnew\ confirmed TNOs then get the same final fit, to all of their data together, with planetary perturbations: the TESS images and the ground-based data, including the ZTF stacks. These are the orbits of Table~\ref{tab:new}. We integrate a test particle under the Sun and eight planets, using the built-in \texttt{astropy} ephemeris \citep{Astropy2022}, with a barycentric state at JD~2460000.5 and topocentric light-time corrections. For each survey with at least four positions, a shared-within-night error component is marginalized in the Gaussian likelihood; its amplitude has a log-uniform prior from $0.001''$ to $3''$. A ZTF stack enters as the TESS images do, with $\ln L=\tfrac12\max(\mathrm{S/N},0)^2$ along the orbit's track as seen from Palomar. A common offset of that track is free, with a Gaussian prior of $0.1''$, just above the $0.045''$--$0.085''$ accuracy of the ZTF astrometric calibration \citep{Masci2019}. The same posterior supplies elements, ephemerides and orbit clones. Two independently initialized ensembles agree in the medians of $a$, $e$ and $i$ within \ensmax\ posterior widths.

We validate predictions by omitting each ground-based night in turn and predicting it from the rest: \loowithin\ of \loonights\ omitted nights fall within $2\sigma$ of the prediction in both sky coordinates (91\% expected for Gaussian errors), with a median residual of $\loomedian''$. This shows that the orbits do not hinge on any single night and that the residuals largely agree with the stated uncertainties. \tno{10} has only one ground night and is excluded. Detailed residuals and the rejected astrometric association are given in Appendix~\ref{sec:orbit-validation}.

For the candidates we have no ground-based positions, and their orbits in Table~\ref{tab:new} are TESS cross-match orbits.

\subsection{Photometry}

The matched-filter flux of each detection converts to $V$ with an offset calibrated on the recovered known objects; the offset scatters by 0.24~mag from object to object, including differences in color. We compute $H_V$ with the distances from the orbit and a phase coefficient of 0.14~mag~deg$^{-1}$, typical of TNOs \citep{Belskaya2003}, and the diameter from $H_V$ and an assumed albedo \citep{PravecHarris2007}. A true diameter needs a measured albedo, so our values are only indicative.

\section{TESS TNO Survey Results}\label{sec:results}
\subsection{New TNOs and candidates}

\begin{figure*}
\centering
\includegraphics[width=\textwidth]{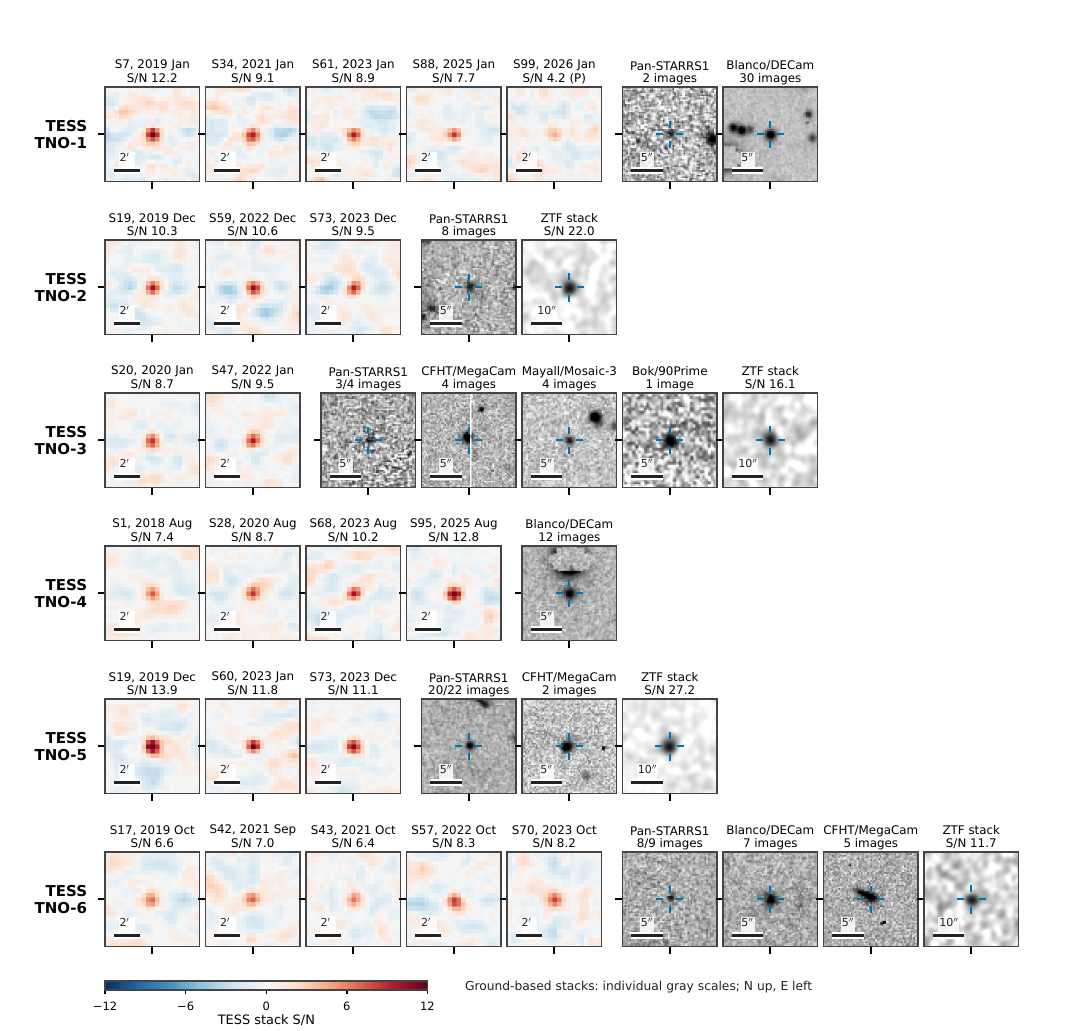}
\caption{Detections of \tno{1} to \tno{6}. TESS panels show each sector stacked along the fitted orbit, with its S/N; (P) marks a recovery on the predicted track instead of a blind recovery. Ground-based panels show the images of each instrument stacked on the object; a fraction gives the images stacked out of the positions measured. The CFHT stack of \tno{3} includes a bad detector column. Figure~\ref{fig:detections-rest} shows \tno{7} to \tno{11}.\label{fig:detections}}
\end{figure*}

The search of \nccd\ CCD fields in \nsec\ sectors yields \nlinked\ linked objects without a designation. Archival ground-based detections confirm \nnew\ of them as new TNOs (Table~\ref{tab:new}; Figures~\ref{fig:orbits} and \ref{fig:detections}; Appendix~\ref{sec:appendix}). Of the others, \ncand\ lack ground-based confirmation and remain candidates, \nruledout\ are excluded (Section~\ref{sec:linking}), and one matches a recently reported Neptune Trojan candidate (Section~\ref{sec:discussion}). The confirmed TNOs span \ranger~au in heliocentric distance and inclinations of \rangei.

\begin{deluxetable*}{llccccccccc}
\tabletypesize{\footnotesize}
\tablecaption{New TNOs and candidates found in TESS images\label{tab:new}}
\tablecolumns{11}
\setlength{\tabcolsep}{3.5pt}
\tablehead{\colhead{Name} & \colhead{Sectors} & \colhead{Arc} & \colhead{$a$} & \colhead{$e$} & \colhead{$i$} & \colhead{$q$} & \colhead{$r$} & \colhead{$V$} & \colhead{$H_V$} & \colhead{$D$} \\
\colhead{} & \colhead{} & \colhead{(yr)} & \colhead{(au)} & \colhead{} & \colhead{(deg)} & \colhead{(au)} & \colhead{(au)} & \colhead{(mag)} & \colhead{(mag)} & \colhead{(km)}}
\startdata
\cutinhead{Confirmed: linked by the cross-match in two or more TESS sectors and detected in archival ground-based images}
TESS TNO-1 & 7, 34, 61, 88 (99) & 13.2 & $58.619(4)$ & $0.36451(3)$ & $21.929865(19)$ & $37.2520(3)$ & 37.6 & 21.7 & 5.9 & 170--430 \\
TESS TNO-2 & 19, 59, 73 & 13.1 & $62.645(20)$ & $0.45106(17)$ & $16.40040(10)$ & $34.388(6)$ & 56.9 & 21.1 & 3.6 & 520--1300 \\
TESS TNO-3 & 20, 47 & 9.1 & $53.993(5)$ & $0.31112(5)$ & $27.13479(5)$ & $37.1951(13)$ & 39.7 & 22.1 & 6.1 & 160--410 \\
TESS TNO-4 & 1, 28, 68, 95 & 10.2 & $41.474(4)$ & $0.01843(7)$ & $22.71127(5)$ & $40.709(6)$ & 42.2 & 21.6 & 5.4 & 230--560 \\
TESS TNO-5 & 19, 60, 73 & 20.3 & $46.1087(14)$ & $0.316839(18)$ & $24.298777(6)$ & $31.49968(15)$ & 31.6 & 21.2 & 6.2 & 150--380 \\
TESS TNO-6 & 17, 42, 43, 57, 70 & 13.1 & $46.4014(25)$ & $0.16518(4)$ & $26.31337(9)$ & $38.7368(12)$ & 42.5 & 22.0 & 5.7 & 190--480 \\
TESS TNO-7 & 13, 67, 105 & 8.0 & $43.695(26)$ & $0.1140(6)$ & $22.74206(10)$ & $38.716(8)$ & 40.5 & 20.8 & 4.7 & 310--770 \\
TESS TNO-8 & 7, 88 & 6.1 & $43.943(10)$ & $0.0994(26)$ & $25.903(5)$ & $39.58(11)$ & 42.2 & 21.6 & 5.4 & 220--560 \\
TESS TNO-9 & 67, 94 & 2.1 & $39.22(7)$ & $0.2394(27)$ & $16.702(3)$ & $29.83(5)$ & 31.2 & 21.7 & 6.7 & 120--300 \\
TESS TNO-10 & 8, 89, 100 & 7.1 & $55.251(9)$ & $0.3034(6)$ & $27.2303(6)$ & $38.49(3)$ & 46.6 & 22.0 & 5.3 & 230--580 \\
TESS TNO-11 & 69, 96, 107 & 9.2 & $39.010(6)$ & $0.36401(8)$ & $21.64038(4)$ & $24.8099(5)$ & 25.0 & 22.1 & 8.0 & 66--170

\\
\cutinhead{Candidates: linked by the cross-match in two or more TESS sectors, not yet detected from the ground}
Candidate-1 & 94, 105 & 1.0 & $64^{+53}_{-14}$ & $0.54 \pm 0.22$ & 18.6 & $29 \pm 4$ & 34.6 & 21.9 & 6.4 & 140--340 \\
Candidate-2 & 94, 95, 105 & 1.0 & $37^{+28}_{-6}$ & $0.46 \pm 0.28$ & 29.1 & $20^{+6}_{-4}$ & 30.4 & 21.9 & 7.0 & 110--260 \\
Candidate-3 & 67, 95, 105 & 3.0 & $173 \pm 23$ & $0.846 \pm 0.019$ & 21.2 & $26.52 \pm 0.08$ & 26.6 & 21.9 & 7.6 & 82--200 \\
Candidate-4 & 67, 95 & 2.1 & $63^{+63}_{-9}$ & $0.47^{+0.31}_{-0.17}$ & 37.2 & $33 \pm 5$ & 38.3 & 22.0 & 6.1 & 160--400 \\
Candidate-5 & 2, 29 & 2.1 & $50^{+25}_{-6}$ & $0.39^{+0.26}_{-0.19}$ & 24.0 & $31 \pm 5$ & 36.9 & 22.2 & 6.5 & 140--340 \\
Candidate-6 & 6, 45 & 3.0 & $59^{+59}_{-10}$ & $0.44 \pm 0.31$ & 12.1 & $33 \pm 7$ & 52.1 & 21.8 & 4.6 & 320--810 \\
Candidate-7 & 94, 105 & 1.0 & $47^{+37}_{-8}$ & $0.45 \pm 0.27$ & 15.5 & $26 \pm 4$ & 31.3 & 21.8 & 6.8 & 120--290 \\
Candidate-8 & 95, 105 & 1.0 & $59^{+66}_{-11}$ & $0.48 \pm 0.30$ & 23.6 & $31 \pm 6$ & 41.0 & 22.1 & 5.9 & 180--440

\enddata
\tablecomments{A sector in parentheses is a recovery on the predicted track instead of the blind search. Elements are barycentric, in the ecliptic J2000 frame. For the confirmed objects they are osculating at JD~2460000.5 (Section~\ref{sec:orbits}), with the 68\% uncertainty of the last digits in parentheses; for the candidates they are medians and 68\% intervals of the orbits that the links allow (within $0.12^\circ$ for $i$). $r$ and $V$ are means over the TESS detections, $H_V$ carries a calibration uncertainty of 0.24~mag, and $D$ assumes albedos of 0.25--0.04.}
\end{deluxetable*}

Among the \ndet\ detections that match no known TNO or Centaur there are \npairs\ possible cross-sector pairs, of which \nlinks\ meet the linking criterion (Section~\ref{sec:linking}). Of these links, \nlinksconf\ join detections of the \nnew\ confirmed TNOs, \nlinkscand\ links join the \ncand\ candidates and \nlinksruled\ links join the \nruledout\ excluded objects; the others belong to Neptune's moon Nereid and to the Trojan candidate.

We confirmed the \nnew\ TNOs by detecting them from the ground, in the archives of Pan-STARRS1, DECam, Mosaic-3, 90Prime, CFHT and ZTF (Table~\ref{tab:ground}; Appendix~\ref{sec:ground}). The first five hold \npositions\ positions of them between \yfirstground\ and \ylastground, and ZTF detects five of them in stacks along their orbits. With the orbits fitted to all of these data together (Section~\ref{sec:orbits}), the positions of the \nnew\ in 2026 November are uncertain by at most $\posmax''$ ($1\sigma$).

\tno{2} is the largest under a common albedo assumption. It has $H_V = \val{2}{H}$ and lies \val{2}{r}~au from the Sun, only $\val{2}{galb}^\circ$ from the Galactic plane at declination $+39^\circ$. Albedos of 0.04--0.25 give a diameter of \val{2}{Dlo}--\val{2}{Dhi}~km \citep{PravecHarris2007}. This illustrative range spans dark to moderately reflective TNO surfaces, as adopted for \edt\ by \citet{Holman2026} and is not a measured size interval. For a representative albedo $p_V=0.08$ \citep{Muller2020}, the diameter is about \val{2}{Dmid}~km. Pan-STARRS1 recorded it on three nights from 2010 to 2014, and a stack of ZTF difference images shows it with an S/N of \val{2}{ztfsnr} (Section~\ref{sec:confirmation}).

The remaining objects span a range of orbital properties and observational arcs (Table~\ref{tab:new}). \tno{4} has a nearly circular orbit; \tno{5} has the longest arc of \val{5}{arc}~years, extended to 2003 by CFHT MegaCam images; and \tno{11}, the nearest and smallest for the adopted albedo, has its perihelion at \val{11}{q}~au, well inside Neptune's orbit. The ground-based evidence is weakest for \tno{10}, which rests on one night of DECam images and a stack of ZTF images. It is in 22 DECam exposures taken over 78~minutes on one night of 2021, moving at the rate that TESS predicts, and in a stack of ZTF images along the orbit through that night (Section~\ref{sec:confirmation}; Appendix~\ref{sec:ztf}).

\begin{figure*}
\centering
\includegraphics[width=\textwidth]{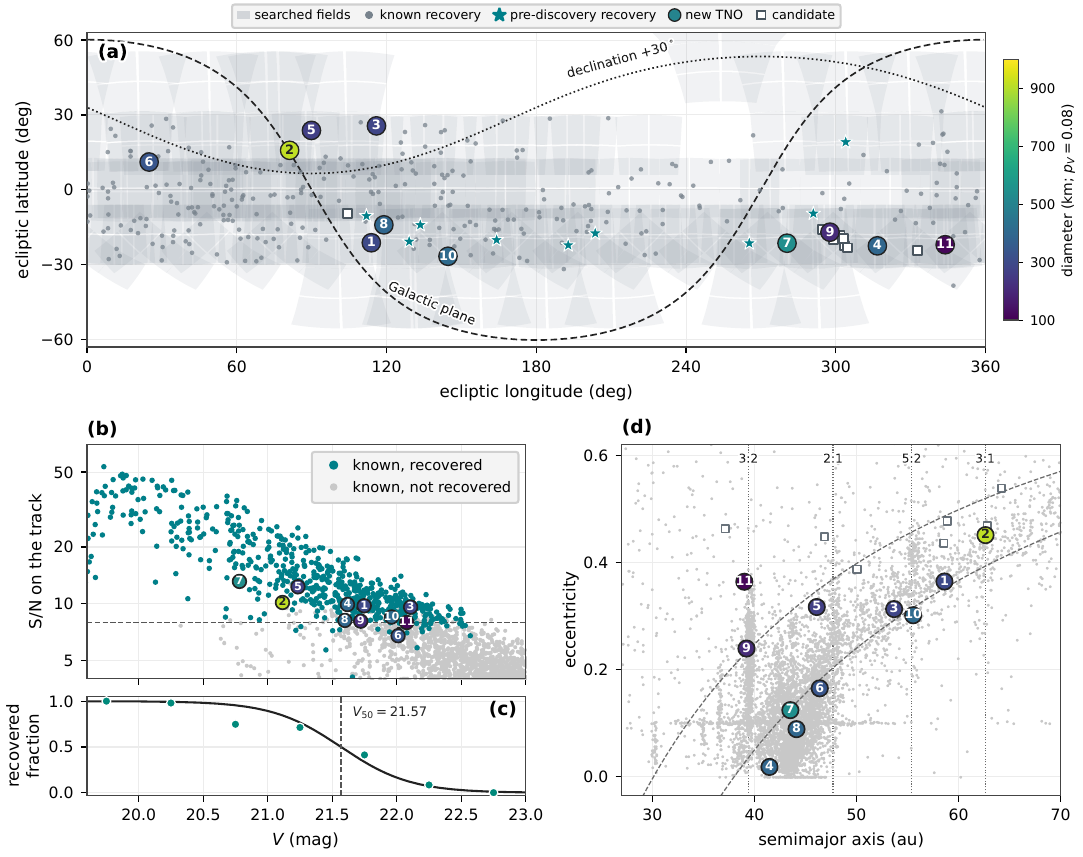}
\caption{TESS TNO Survey coverage and recovery. (a) Searched fields, new TNOs (numbered, colored by diameter for $p_V=0.08$), candidates (open squares) and pre-discovery recoveries (stars). (b) S/N on the track versus $V$ for known TNOs and for the new TNOs (numbered, at the mean of their sectors); the dashed line marks the recovery limit of 7.95 in local S/N. (c) Recovered fraction with a logistic fit. (d) Eccentricity versus semimajor axis, with Neptune's resonances and perihelia of 30 and 38~au.\label{fig:survey}}
\end{figure*}

None of the \nnew\ TNOs is a designated object in the JPL Small-Body Database. We compared each of them with the JPL small-body predictions within $0.5^\circ$ of its TESS detections, down to $V=26$, and their final orbits with 9{,}423 cataloged TNO and Centaur orbits (2026 November 10), using orbital plane, heliocentric distance and position. The only case passing the broad plane, distance and position filters, 2014~RS$_{86}$, lies $4.2^\circ$ from \tno{4} on a different orbit. The same positional queries recover known objects within $6''$--$38''$ of their TESS detections. We find no earlier report of the \nnew\ TNOs.

The \ncand\ candidates have blind detections in two or three TESS sectors, with local S/N of \candlocal, but no secure independent ground-based archival recovery. Four lie in crowded fields with poorly constrained earlier positions; the others have too few possible ground detections to establish an association. Their orbital intervals in Table~\ref{tab:new} are conditional on the TESS links being real. 

\subsection{Recovery of known TNOs in TESS images}\label{sec:known}

JPL Horizons \citep{Giorgini1996} predicts \nappear\ appearances of known TNOs and Centaurs within our searched velocity range. The blind search recovers \nknown\ objects (\nknowntno\ TNOs and \nknowncen\ Centaurs), including Makemake, Haumea, Quaoar, Orcus, and objects as distant as Eris at 96~au; \nknowntwo\ appear in multiple sectors. It recovers \nbrightrec\ of \nbright\ TNO appearances brighter than $V=20.5$, falling to 50\% at $V=\vhalf$ (Figure~\ref{fig:survey}c).

For this comparison, recovery requires a blind detection on the JPL track with local S/N~$\geq7.95$, the level above which the control search of our first 18 sectors left no detection; we keep it only as a fixed definition of recovery. Of \nknowndet\ such detections, \nknownpass\ (\knownpasspct\%) pass the quality cuts. The TESS cross-match links \nknownlinks\ of \nknownpairs\ same-object pairs (\knownlinkpct\%). 

\Npre\ of the recovered objects were designated only after TESS imaged them. They include \edt, which the search detects in Sector~12 (2019 June), 6.8~years before its discovery \citep{Holman2026}, and again in 2021, 2023 and 2025 \citep{Han2026}. Several recently designated objects are recovered more strongly on the blind-search TESS track than on the JPL prediction, illustrating the value of TESS precovery for short-arc orbits.

\section{Discussion}\label{sec:discussion}

The new TNOs occupy regions poorly covered by many deep surveys: TESS observed them \rangelat\ from the ecliptic, and several lie in northern or crowded fields (Figure~\ref{fig:survey}a; \citealt{Petit2011,Bannister2018,Bernardinelli2022}). Their locations plausibly explain why they remained unidentified. TGLC star modeling is particularly useful near the Galactic plane and in other crowded fields, where the residuals of unsubtracted stars would otherwise dominate the stacks.

All \nnew\ TESS TNOs were nevertheless already in ground-based archives, most of them on several nights. TESS links between sectors provide the orbital constraints needed to associate these scattered observations, which then improve astrometric precision significantly over TESS alone. The \ncand\ candidates lack comparable independent support in the archives examined and demand further targeted ground observations. 

TESS and Rubin can be complementary in the search for TNOs. Rubin reaches several magnitudes deeper with much better resolution, but TESS covers the northern sky that Rubin cannot reach. For the objects Rubin does find, the TESS archive already holds observations from as early as 2018 for immediate follow-up, as the precoveries in Section~\ref{sec:known} show.

Our TNO search in TESS can be taken further with injection--recovery tests. Known-object recovery measures an empirical sensitivity; the full discovery completeness also depends on the quality cuts, the linking and the archival coverage. Injecting synthetic objects before cleaning and recovering them through the whole pipeline would quantify these losses, especially for slow sources, and would let us turn the detections into an occurrence rate for bright TNOs at high ecliptic latitude. Until then we make no population estimate for undiscovered objects.

The TESS cross-match also links a recently reported Neptune Trojan candidate. Our measured orbit ($a=\tval{aaa}$~au, $e=\tval{ee}$, $i=\tval{i}^\circ$) closely resembles candidate GCT0044, reported by \citet{Cerati2026}. The identity remains provisional because that report withholds positions. It is also detected from the ground: Pan-STARRS1, DECam and 90Prime give \tval{npos} positions on \tval{nights} nights between \tval{yfirst} and \tval{ylast}, and a ZTF stack shows it with an S/N of \tval{ztfsnr}. In a 1~Myr \texttt{REBOUND} integration \citep{ReinLiu2012} with the four giant planets, all \troclones\ posterior clones librate around L4, with amplitude $\troamp^{+\troamphi}_{-\troamplo}$~deg and period \troperiod~kyr. This establishes libration over the tested interval only.

TESS and archival ground imaging together provide a practical route to identifying bright outer Solar System objects and recovering their earlier positions.

\begin{acknowledgments}
This paper includes data collected by the TESS mission, funded by NASA's Science Mission Directorate, and data from the ESA mission Gaia, processed by the Gaia Data Processing and Analysis Consortium (DPAC), funded by national institutions, in particular those in the Gaia Multilateral Agreement. We thank the observing, processing and archive teams whose work made these archival measurements possible.

The NOIRLab Source Catalog, Astro Data Lab and Astro Data Archive supplied catalogs and images for this work. AURA operates NSF NOIRLab under an agreement with the U.S. National Science Foundation. DECam was built by the Dark Energy Survey collaboration, with support from the U.S. Department of Energy, NSF, international funding agencies and the participating institutions. We used Pan-STARRS1 survey products served by STScI through MAST; the PS1 consortium, NASA, NSF and the Gordon and Betty Moore Foundation supported the surveys and their public archive.

The CFHT images were obtained with MegaPrime/MegaCam, developed jointly by CFHT and CEA. CFHT is operated jointly by Canada's National Research Council, France's CNRS/INSU and the University of Hawaii. Access was provided by the Canadian Astronomy Data Centre, run by the National Research Council of Canada with Canadian Space Agency support. The ZTF observations were made with Palomar's Samuel Oschin telescope; NSF and the ZTF partner institutions support the project, with operations carried out by Caltech Optical Observatories, IPAC and the University of Washington. We retrieved ZTF products through IRSA, the NASA-funded archive operated at Caltech. We also used the JPL Small-Body Database and Horizons.

We used Claude Opus 5.5 and Fable 5.2 (Anthropic), and GPT-6 Astra through Codex (OpenAI), for assistance with analysis code, figures and manuscript preparation; the authors reviewed the results and take responsibility for the content.
\end{acknowledgments}

\facilities{TESS, Gaia, PO:1.2m (ZTF), Blanco (DECam), Mayall (Mosaic-3), Bok (90Prime), PS1, CFHT (MegaCam), MAST, IRSA, CADC, Astro Data Lab}

\software{astropy \citep{Astropy2022}, emcee \citep{ForemanMackey2013}, REBOUND \citep{ReinLiu2012}, TGLC \citep{Han2023}, tess-point \citep{Burke2020}, SEP \citep{Barbary2016,BertinArnouts1996}, NumPy \citep{Harris2020}, SciPy \citep{Virtanen2020}, CuPy \citep{Okuta2017}, Matplotlib \citep{Hunter2007}}

\appendix

\section{Detections}\label{sec:appendix}

\begin{figure*}[!htbp]
\centering
\includegraphics[width=\textwidth]{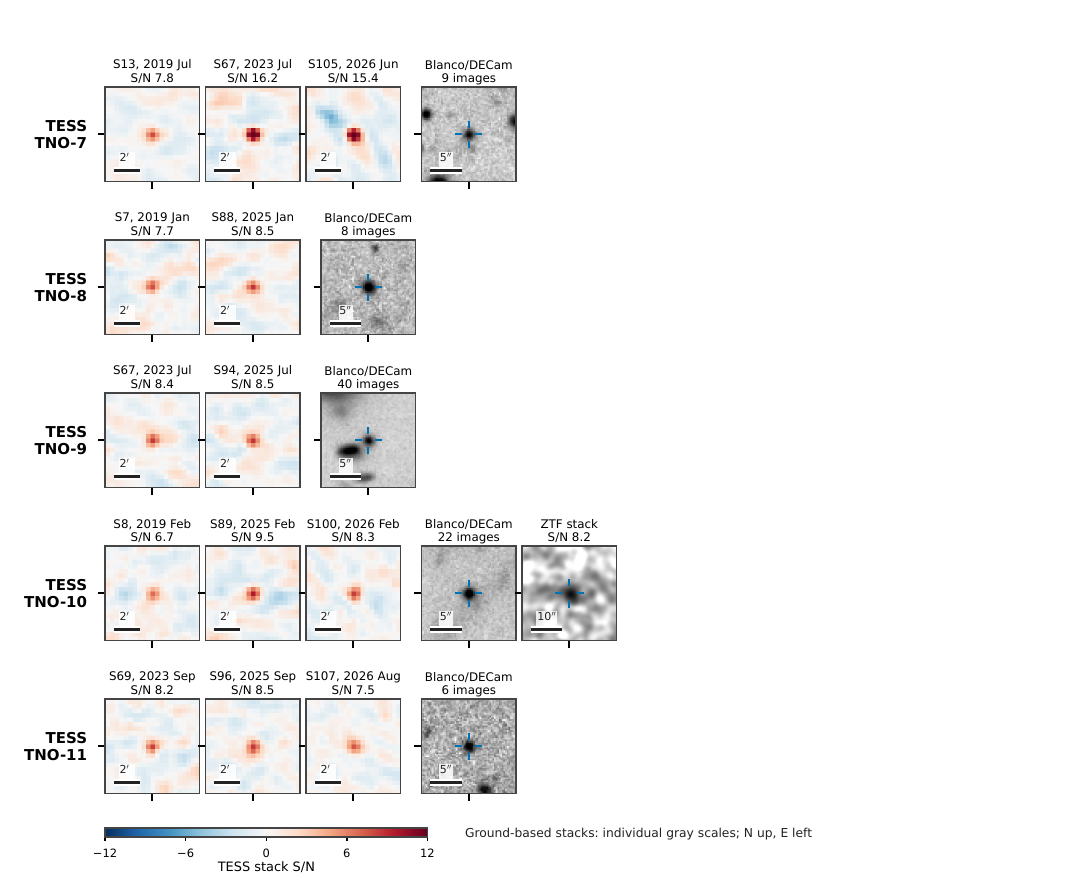}
\caption{Detections of \tno{7} to \tno{11}, in the format of Figure~\ref{fig:detections}.\label{fig:detections-rest}}
\end{figure*}

Figure~\ref{fig:detections-rest} continues Figure~\ref{fig:detections}. Tables~\ref{tab:detections} and \ref{tab:candidates} list every TESS detection of the \nnew\ TNOs and of the candidates.

\begin{deluxetable*}{llllccccrr}
\tabletypesize{\footnotesize}
\tablecaption{TESS detections of the new TNOs\label{tab:detections}}
\tablecolumns{10}
\tablehead{\colhead{Name} & \colhead{Sector} & \colhead{Dates (UT)} & \colhead{Camera-CCD} & \colhead{Found} & \colhead{Local S/N} & \colhead{S/N on} & \colhead{Flux} & \colhead{$r$} & \colhead{$\Delta$} \\
\colhead{} & \colhead{} & \colhead{} & \colhead{} & \colhead{} & \colhead{(blind)} & \colhead{the orbit} & \colhead{(e$^-$\,s$^{-1}$)} & \colhead{(au)} & \colhead{(au)}}
\startdata
TESS TNO-1 & 7 & 2019 Jan 8--31 & 1-1 & blind & 11.8 & 12.2 & $0.62 \pm 0.05$ & 37.3 & 36.4 \\
 & 34 & 2021 Jan 14--Feb 8 & 1-1 & blind & 7.9 & 9.1 & $0.47 \pm 0.05$ & 37.4 & 36.5 \\
 & 61 & 2023 Jan 19--Feb 12 & 1-1 & blind & 8.4 & 8.9 & $0.45 \pm 0.05$ & 37.6 & 36.7 \\
 & 88 & 2025 Jan 17--Feb 10 & 1-1 & blind & 9.3 & 7.7 & $0.54 \pm 0.07$ & 37.8 & 36.9 \\
 & 99 & 2026 Jan 9--30 & 1-4 & predicted & -- & 4.2 & $0.28 \pm 0.07$ & 37.9 & 37.0 \\
TESS TNO-2 & 19 & 2019 Nov 28--Dec 22 & 1-4 & blind & 10.1 & 10.3 & $0.81 \pm 0.08$ & 57.8 & 56.9 \\
 & 59 & 2022 Nov 27--Dec 23 & 1-4 & blind & 9.7 & 10.6 & $0.88 \pm 0.08$ & 56.7 & 55.7 \\
 & 73 & 2023 Dec 7--2024 Jan 3 & 1-3 & blind & 8.8 & 9.5 & $1.02 \pm 0.11$ & 56.3 & 55.3 \\
TESS TNO-3 & 20 & 2019 Dec 25--2020 Jan 20 & 1-1 & blind & 8.9 & 8.7 & $0.33 \pm 0.04$ & 39.5 & 38.6 \\
 & 47 & 2022 Jan 2--27 & 1-1 & blind & 9.3 & 9.5 & $0.39 \pm 0.04$ & 39.9 & 39.0 \\
TESS TNO-4 & 1 & 2018 Jul 25--Aug 22 & 1-2 & blind & 8.9 & 7.4 & $0.55 \pm 0.07$ & 42.2 & 41.3 \\
 & 28 & 2020 Jul 31--Aug 21 & 1-1 & blind & 9.1 & 8.7 & $0.59 \pm 0.07$ & 42.2 & 41.3 \\
 & 68 & 2023 Jul 29--Aug 21 & 1-2 & blind & 9.3 & 10.2 & $0.44 \pm 0.04$ & 42.2 & 41.3 \\
 & 95 & 2025 Jul 27--Aug 19 & 1-2 & blind & 11.6 & 12.8 & $0.61 \pm 0.05$ & 42.2 & 41.3 \\
TESS TNO-5 & 19 & 2019 Nov 28--Dec 22 & 1-1 & blind & 14.5 & 13.9 & $0.83 \pm 0.06$ & 31.5 & 30.6 \\
 & 60 & 2022 Dec 24--2023 Jan 18 & 1-2 & blind & 12.1 & 11.8 & $0.80 \pm 0.07$ & 31.6 & 30.7 \\
 & 73 & 2023 Dec 7--2024 Jan 3 & 1-1 & blind & 11.9 & 11.1 & $0.77 \pm 0.07$ & 31.7 & 30.8 \\
TESS TNO-6 & 17 & 2019 Oct 8--29 & 1-3 & blind & 6.7 & 6.6 & $0.47 \pm 0.07$ & 42.8 & 41.9 \\
 & 42 & 2021 Aug 21--Sep 12 & 3-4 & blind & 7.4 & 7.0 & $0.39 \pm 0.06$ & 42.6 & 41.9 \\
 & 43 & 2021 Sep 22--Oct 10 & 1-2 & blind & 8.2 & 6.4 & $0.40 \pm 0.06$ & 42.6 & 41.6 \\
 & 57 & 2022 Sep 30--Oct 29 & 1-4 & blind & 10.8 & 8.3 & $0.33 \pm 0.04$ & 42.4 & 41.4 \\
 & 70 & 2023 Sep 21--Oct 15 & 3-1 & blind & 8.9 & 8.2 & $0.39 \pm 0.05$ & 42.3 & 41.4 \\
TESS TNO-7 & 13 & 2019 Jul 4--17 & 1-1 & blind & 11.1 & 7.8 & $1.48 \pm 0.19$ & 40.2 & 39.2 \\
 & 67 & 2023 Jul 1--25 & 1-1 & blind & 15.4 & 16.2 & $1.33 \pm 0.08$ & 40.5 & 39.5 \\
 & 105 & 2026 Jun 16--Jul 11 & 1-4 & blind & 13.5 & 15.4 & $1.05 \pm 0.07$ & 40.8 & 39.8 \\
TESS TNO-8 & 7 & 2019 Jan 8--25 & 1-4 & blind & 8.5 & 7.7 & $0.58 \pm 0.07$ & 42.4 & 41.5 \\
 & 88 & 2025 Jan 17--Feb 10 & 1-4 & blind & 6.9 & 8.5 & $0.54 \pm 0.06$ & 41.9 & 41.0 \\
TESS TNO-9 & 67 & 2023 Jul 1--25 & 1-3 & blind & 8.9 & 8.4 & $0.57 \pm 0.07$ & 31.0 & 30.0 \\
 & 94 & 2025 Jun 30--Jul 24 & 1-3 & blind & 9.8 & 8.5 & $0.47 \pm 0.06$ & 31.3 & 30.4 \\
TESS TNO-10 & 8 & 2019 Feb 4--27 & 1-1 & blind & 7.2 & 6.7 & $0.36 \pm 0.05$ & 47.7 & 46.9 \\
 & 89 & 2025 Feb 13--Mar 10 & 1-1 & blind & 8.2 & 9.5 & $0.44 \pm 0.05$ & 46.2 & 45.3 \\
 & 100 & 2026 Feb 5--26 & 1-4 & blind & 8.7 & 8.3 & $0.39 \pm 0.05$ & 45.9 & 45.0 \\
TESS TNO-11 & 69 & 2023 Aug 26--Sep 17 & 1-1 & blind & 8.5 & 8.2 & $0.40 \pm 0.05$ & 24.9 & 23.9 \\
 & 96 & 2025 Aug 22--Sep 14 & 1-2 & blind & 9.3 & 8.5 & $0.35 \pm 0.04$ & 25.0 & 24.1 \\
 & 107 & 2026 Aug 9--Sep 7 & 1-3 & blind & 8.5 & 7.5 & $0.38 \pm 0.05$ & 25.1 & 24.3

\enddata
\tablecomments{``Predicted'' means a recovery on the track from the other sectors. $r$ and $\Delta$ are the distances from the Sun and from TESS at mid-sector. Scattered light flagged and removed 37\% of the Sector~99 images.}
\end{deluxetable*}

\begin{deluxetable*}{lllcccrr}
\tabletypesize{\footnotesize}
\tablecaption{TESS detections of the candidates\label{tab:candidates}}
\tablecolumns{8}
\tablehead{\colhead{Name} & \colhead{Sector} & \colhead{Dates (UT)} & \colhead{Camera-CCD} & \colhead{Local S/N} & \colhead{Flux} & \colhead{$r$} & \colhead{$\Delta$} \\
\colhead{} & \colhead{} & \colhead{} & \colhead{} & \colhead{(blind)} & \colhead{(e$^-$\,s$^{-1}$)} & \colhead{(au)} & \colhead{(au)}}
\startdata
Candidate-1 & 94 & 2025 Jun 30--Jul 24 & 1-3 & 9.7 & $0.49 \pm 0.05$ & 34.7 & 33.7 \\
 & 105 & 2026 Jun 15--Jul 11 & 1-3 & 8.5 & $0.41 \pm 0.05$ & 34.5 & 33.6 \\
Candidate-2 & 94 & 2025 Jun 30--Jul 24 & 1-2 & 9.5 & $0.43 \pm 0.05$ & 30.4 & 29.5 \\
 & 95 & 2025 Jul 27--Aug 19 & 1-1 & 7.2 & $0.41 \pm 0.06$ & 30.4 & 29.5 \\
 & 105 & 2026 Jun 16--Jul 11 & 1-2 & 8.8 & $0.45 \pm 0.05$ & 30.2 & 29.4 \\
Candidate-3 & 67 & 2023 Jul 1--25 & 1-2 & 7.6 & $0.46 \pm 0.06$ & 26.6 & 25.7 \\
 & 95 & 2025 Jul 27--Aug 19 & 1-1 & 9.2 & $0.47 \pm 0.05$ & 26.5 & 25.6 \\
 & 105 & 2026 Jun 16--Jul 11 & 1-2 & 9.4 & $0.40 \pm 0.04$ & 26.5 & 25.8 \\
Candidate-4 & 67 & 2023 Jul 1--25 & 1-2 & 7.5 & $0.43 \pm 0.06$ & 37.9 & 37.0 \\
 & 95 & 2025 Jul 27--Aug 19 & 1-4 & 9.1 & $0.40 \pm 0.04$ & 38.7 & 37.7 \\
Candidate-5 & 2 & 2018 Aug 23--Sep 17 & 1-1 & 7.9 & $0.33 \pm 0.04$ & 37.1 & 36.2 \\
 & 29 & 2020 Aug 26--Sep 17 & 1-1 & 7.8 & $0.34 \pm 0.04$ & 36.8 & 35.9 \\
Candidate-6 & 6 & 2018 Dec 17--2019 Jan 6 & 1-3 & 7.7 & $0.53 \pm 0.07$ & 52.2 & 51.2 \\
 & 45 & 2021 Nov 8--Dec 2 & 2-4 & 6.6 & $0.43 \pm 0.07$ & 52.0 & 51.4 \\
Candidate-7 & 94 & 2025 Jun 30--Jul 24 & 1-3 & 7.0 & $0.44 \pm 0.06$ & 31.1 & 30.1 \\
 & 105 & 2026 Jun 15--Jul 11 & 1-3 & 7.6 & $0.51 \pm 0.07$ & 31.5 & 30.6 \\
Candidate-8 & 95 & 2025 Jul 27--Aug 19 & 1-1 & 7.5 & $0.36 \pm 0.05$ & 41.1 & 40.2 \\
 & 105 & 2026 Jun 16--Jul 11 & 1-2 & 7.2 & $0.37 \pm 0.05$ & 40.9 & 40.1

\enddata
\tablecomments{All detections are blind. Fluxes are along the search path; $r$ and $\Delta$ are from the linked orbit.}
\end{deluxetable*}

\section{Survey diagnostics}\label{sec:diagnostics}

We use the known objects to calibrate this survey. Of \nknownpairs\ pairs of detections of the same known object, \nknownlinks\ pass the linking criterion, so the TESS cross-match links \knownlinkpct\% of the same-object pairs detected in two sectors. In the others the measured motions differ from those of the orbit by more than our tolerance. We also measure the rate of chance links, using pairs that cannot be real. Of \ncontrolpairs\ pairs in which one detection comes from the control search, \ncontrollinks\ are links, which scales to \nchance\ among the \npairs\ pairs of new detections. Every link in this control sample includes a weak detection. Where the weaker detection of the pair has a local S/N below 6.5, the control predicts \nchancelow\ links and we find none among the new detections; between 6.5 and 7.2 it predicts \nchancemid\ and we find \nlinksmid; above 7.2, where \nlinksstrong\ of our \nlinks\ links lie, not one of the \ncontrolpairs\ control pairs is a link. The control's links join two detections each, except for one family of five detections in two fields and one control detection that links to two real detections of \edt. One pair of new detections meets every condition except the inclination: its orbit is inclined by $85^\circ$, and it is the weakest pair of all (local S/N 6.7 and 5.8). The same is true of \ncontrolhighi\ pairs of the control.

\section{Orbit validation}\label{sec:orbit-validation}

As a test of the orbits and of their uncertainties, we refit each orbit without each of its ground-based nights in turn and predict that night (\loonights\ nights of the ten TNOs that have more than one). The omitted night lies a median of $\loomedian''$ from its prediction, and within twice the local predicted uncertainty in both sky coordinates for \loowithin\ of the \loonights. The largest deviations are $\val{2}{loomax}\sigma$ for the 2010 night of \tno{2}, $\val{5}{loomax}\sigma$ for one Pan-STARRS1 night of \tno{5}, $\val{4}{loomax}\sigma$ for \tno{4} and $\val{9}{loomax}\sigma$ for \tno{9}. \tno{2} and \tno{9} have only three ground-based nights, so that leaving one out leaves little, and their omitted nights miss by up to $\val{2}{loolargest}''$ and $\val{9}{loolargest}''$. The 2010 night of \tno{2} misses along the track and agrees to $\val{2}{looacross}''$ across it; when we sample the posterior without that night, it lies $\val{2}{loosampled}\sigma$ from the prediction. For \tno{10}, a refit to the TESS and ZTF images alone misses the sole ground night by $\val{10}{loolargest}''$. This is an observed residual; the formal uncertainty of that orbit is not validated by this local approximation. Its independent DECam motion and ZTF detection support the association, but its single ground night provides weaker validation than the multiple-night cases. The test also rejected one position. A DECam source of 2024 May that we had first attributed to \tno{7} lies $10''$ ($11\sigma$) from where the rest of the data put the object, and with it the orbit missed the object in Sector~105 by 1.4~pixels; without it, one orbit fits the remaining \val{7}{nnum} coordinates ($\chi^2 = \val{7}{chipert}$) and all three TESS sectors, and we exclude it. On the final orbits the \nztffit\ ZTF stacks have S/N of \ztffitmin--\ztffitmax, against \ztfsearchmin--\ztfsearchmax\ at the maxima of their searches (Appendix~\ref{sec:ztf}), and the fitted offsets of their tracks are at most $\ztfoffmax''$.

For the omitted-night test we fix the error-component amplitudes at their full-fit medians and approximate the refitted covariance locally. This diagnostic can underrepresent weak or non-Gaussian directions; it does not replace posterior sampling of each omission.

\section{Archival detections from the ground}\label{sec:ground}\label{sec:ztf}

Table~\ref{tab:ground} lists the ground-based observations. For NSC positions we add $0.05''$ in quadrature to the catalog uncertainty. For the others the centroid uncertainty is the seeing FWHM over 2.35 times the S/N (assuming $1.2''$ seeing for Pan-STARRS1), plus $0.06''$ in quadrature for Pan-STARRS1 and the reference-star scatter for CFHT, and at least $0.08''$ for DECam. The uncertainties range from $0.07''$ to $0.28''$.

The weakest cases are the following. \tno{2} has Pan-STARRS1 positions on only three nights, $5^\circ$ from the Galactic plane; ZTF also detects it, and its orbit, fitted without the images of Sector~73, put it within a pixel of where the blind search found it there. \tno{7} is only in DECam images of a crowded field, on three nights in 2018 and two in 2021; a position on a sixth night, in 2024, failed the omitted-night test (Appendix~\ref{sec:orbit-validation}) and is not used. \tno{8} is seen in one DECam season of 55 days. \tno{9}, \tno{10} and \tno{11} were recovered last, from orbits that only the TESS cross-match constrained, for which we take the TESS positions to be uncertain by one pixel. \tno{9} is on three nights in 2025, the second predicted from the other two to $0.3''$ and found $0.27''$ away; \tno{11} is on six nights in 2017 and 2023, one predicted from two others to $1.9'' \times 1.0''$ and found $0.31''$ away; \tno{10} is on one night. Orbits fitted to these positions with only the cross-match as the TESS constraint give the TESS images of \tno{9} and \tno{11} an S/N of 7.5--8.5 in every sector, as in the final fits (Table~\ref{tab:detections}), and those of \tno{10} 4.5--6.9.

For the ZTF stacks we filter the public difference images with a Gaussian matched to their seeing, estimate the noise from the image scatter, mask strong subtraction residuals, and combine the fluxes with inverse-variance weights. We maximize the S/N over 1500 posterior orbits and offsets of up to $6''$ in each coordinate, and repeat the search at 192 displaced locations $25''$--$140''$ away; a detection must exceed every one of its displaced searches and an S/N of 8, about the highest value that the displaced searches of the detected objects reach (their maxima are 5.3--8.1). The stacks of \tno{2}, \tno{3}, \tno{5} and \tno{6} use $r$-band images near a TESS observing season and survive the removal of any single night; adding their likelihood to that of the TESS images narrowed the orbits with which we searched the other archives. For \tno{10} the trial orbits pass through its DECam night, and its 70 usable images (42 in $r$, 26 in $g$, two in $i$) give an S/N of 8.2, the $r$ images alone 7.9; removing any one night leaves at least 7.7. The final orbit fits use the exposures of these stacks (Section~\ref{sec:orbits}). The three excluded objects reach S/N 5.1, 8.5 and 5.8 in such mixed-filter stacks, below the maxima of their displaced searches (5.9, 9.7 and 7.9), where their TESS brightnesses predict 28, 19 and 17. Colors, variability and incomplete orbital coverage can affect such a nonrecovery.

\begin{deluxetable*}{lllccllc}
\tabletypesize{\footnotesize}
\tablecaption{Ground-based observations of the new TNOs\label{tab:ground}}
\tablecolumns{8}
\tablehead{\colhead{Name} & \colhead{Telescope} & \colhead{Camera} & \colhead{Positions} & \colhead{Nights} & \colhead{Dates (UT)} & \colhead{Filters} & \colhead{Magnitude}}
\startdata
TESS TNO-1 & Pan-STARRS1 1.8 m & Pan-STARRS1 & 2 & 1 & 2012 Nov 7 & $i$ & 21.5--21.6 \\
 & Blanco 4 m & DECam & 30 & 7 & 2015 Mar 7--2024 Mar 5 & $r$, $i$, $z$, VR & 21.5--22.1 \\
TESS TNO-2 & Pan-STARRS1 1.8 m & Pan-STARRS1 & 8 & 3 & 2010 Nov 17--2014 Dec 3 & $g$, $r$, $i$ & 20.8--22.1 \\
 & Palomar 1.2 m & ZTF & stack of 38 & 29 & 2019 Nov 3--2020 Jan 31 & $r$ & 21.45 \\
TESS TNO-3 & Pan-STARRS1 1.8 m & Pan-STARRS1 & 4 & 2 & 2012 Dec 31--2015 Jan 11 & $i$ & 21.3--21.5 \\
 & Mayall 4 m & Mosaic-3 & 4 & 4 & 2016 Mar 11--2017 Mar 11 & $z$ & 21.4--21.8 \\
 & CFHT 3.6 m & MegaCam & 4 & 3 & 2017 Dec 23--26 & $r$ & 21.9--22.4 \\
 & Bok 2.3 m & 90Prime & 1 & 1 & 2018 Jan 18 & $r$ & 21.7 \\
 & Palomar 1.2 m & ZTF & stack of 107 & 59 & 2019 Oct 10--2020 Mar 7 & $r$ & 22.43 \\
TESS TNO-4 & Blanco 4 m & DECam & 12 & 7 & 2015 Jun 6--2020 Nov 25 & $g$, $r$, $i$, $z$ & 20.8--22.7 \\
TESS TNO-5 & CFHT 3.6 m & MegaCam & 2 & 1 & 2003 Sep 21 & $r$ & 21.3--21.7 \\
 & Pan-STARRS1 1.8 m & Pan-STARRS1 & 22 & 10 & 2010 Nov 18--2013 Dec 25 & $g$, $r$, $i$, $z$ & 20.7--22.2 \\
 & Palomar 1.2 m & ZTF & stack of 80 & 50 & 2019 Sep 14--2020 Mar 7 & $r$ & 21.56 \\
TESS TNO-6 & Pan-STARRS1 1.8 m & Pan-STARRS1 & 9 & 4 & 2010 Sep 17--2014 Dec 8 & $r$, $i$ & 21.3--22.4 \\
 & Blanco 4 m & DECam & 7 & 7 & 2014 Aug 11--2018 Oct 17 & $g$, $r$, $z$ & 21.3--23.1 \\
 & CFHT 3.6 m & MegaCam & 5 & 2 & 2016 Oct 22--2018 Jan 16 & $gri$ & 22.0--22.3 \\
 & Palomar 1.2 m & ZTF & stack of 23 & 21 & 2021 Jul 7--Nov 22 & $r$ & 22.27 \\
TESS TNO-7 & Blanco 4 m & DECam & 9 & 5 & 2018 Jul 21--2021 Apr 26 & $r$, $i$, $Y$ & 20.6--20.9 \\
TESS TNO-8 & Blanco 4 m & DECam & 8 & 7 & 2019 Dec 11--2020 Feb 4 & $g$, $i$, VR & 21.5--22.2 \\
TESS TNO-9 & Blanco 4 m & DECam & 40 & 3 & 2025 Apr 3--Jun 2 & $g$, $r$, $i$, $z$ & 21.7--22.6 \\
TESS TNO-10 & Blanco 4 m & DECam & 22 & 1 & 2021 Apr 13 & $r$ & 22.0--22.2 \\
 & Palomar 1.2 m & ZTF & stack of 70 & 60 & 2021 May 1--2025 Dec 21 & $g$, $r$, $i$ & \nodata \\
TESS TNO-11 & Blanco 4 m & DECam & 6 & 6 & 2017 Jul 2--2023 Aug 27 & $r$, $z$ & 21.7--22.2

\enddata
\tablecomments{Pan-STARRS1, Mosaic-3 and 90Prime positions are cataloged detections, MegaCam positions are measured on images, and DECam positions are of both kinds. Magnitudes are in the filters listed, not in $V$; $gri$ is the wide MegaCam filter. ZTF rows are stacks; the mixed-filter stack of \tno{10} has no magnitude.}
\end{deluxetable*}

\bibliography{tess_tno}{}

@article{Han2023,
  author = {{Han}, Te and {Brandt}, Timothy D.},
  title = {TESS-Gaia Light Curve: A PSF-based TESS FFI Light-curve Product},
  journal = {AJ},
  year = 2023,
  volume = 165,
  pages = {71},
  doi = {10.3847/1538-3881/acaaa7}
}

@article{Han2026,
  author = {{Han}, Te and {Audenaert}, Jeroen and {Muthukrishna}, Daniel and {Holman}, Matthew and {Ricker}, George R.},
  title = {Seven Years of TESS Precovery Observations of the Dwarf-planet Candidate 2026 ED$_{138}$},
  journal = {RNAAS},
  year = 2026,
  volume = 10,
  pages = {332},
  doi = {10.3847/2515-5172/aec006},
  eprint = {2610.40106},
  archivePrefix = {arXiv}
}

@article{Ricker2015,
  author = {{Ricker}, George R. and {Winn}, Joshua N. and {Vanderspek}, Roland and others},
  title = {Transiting Exoplanet Survey Satellite (TESS)},
  journal = {JATIS},
  year = 2015,
  volume = 1,
  pages = {014003},
  doi = {10.1117/1.JATIS.1.1.014003}
}

@article{Fausnaugh2020,
  author = {{Fausnaugh}, Michael M. and {Burke}, Christopher J. and {Ricker}, George R. and {Vanderspek}, Roland},
  title = {Calibrated Full-frame Images for the TESS Quick Look Pipeline},
  journal = {RNAAS},
  year = 2020,
  volume = 4,
  pages = {251},
  doi = {10.3847/2515-5172/abd63a}
}

@article{Gaia2016,
  author = {{\unskip Gaia Collaboration} and {Prusti}, T. and others},
  title = {The Gaia mission},
  journal = {A\&A},
  year = 2016,
  volume = 595,
  pages = {A1},
  doi = {10.1051/0004-6361/201629272}
}

@article{Gaia2023,
  author = {{\unskip Gaia Collaboration} and {Vallenari}, A. and others},
  title = {Gaia Data Release 3. Summary of the content and survey properties},
  journal = {A\&A},
  year = 2023,
  volume = 674,
  pages = {A1},
  doi = {10.1051/0004-6361/202243940}
}

@article{Giorgini1996,
  author = {{Giorgini}, J. D. and {Yeomans}, D. K. and {Chamberlin}, A. B. and others},
  title = {JPL's On-Line Solar System Data Service},
  journal = {BAAS},
  year = 1996,
  volume = 28,
  pages = {1158}
}

@article{Ivezic2019,
  author = {{Ivezi{\'c}}, {\v Z}eljko and {Kahn}, Steven M. and {Tyson}, J. Anthony and others},
  title = {LSST: From Science Drivers to Reference Design and Anticipated Data Products},
  journal = {ApJ},
  year = 2019,
  volume = 873,
  pages = {111},
  doi = {10.3847/1538-4357/ab042c}
}

@ARTICLE{MPEC2026U85,
  author = {{Juri{\'c}}, M. and {Holman}, M.~J. and {Bernardinelli}, P.~H. and {Eggl}, S. and {Heinze}, A. and {Kurlander}, J. and {Moeyens}, J. and {Napier}, K. and {Nourbakhsh}, E. and {Alexandersen}, M.},
  title = "{2026 ED138}",
  journal = {Minor Planet Electronic Circulars},
  year = 2026, month = oct,
  volume = {2026-U85},
  url = {https://www.minorplanetcenter.net/mpec/K26/K26U85.html}
}

@article{RiceLaughlin2020,
  author = {{Rice}, Malena and {Laughlin}, Gregory},
  title = {Exploring Trans-Neptunian Space with TESS: A Targeted Shift-stacking Search for Planet Nine and Distant TNOs in the Galactic Plane},
  journal = {PSJ},
  year = 2020,
  volume = 1,
  pages = {81},
  doi = {10.3847/PSJ/abc42c}
}

@article{delaFuenteMarcos2022,
  author = {{de la Fuente Marcos}, C. and {de la Fuente Marcos}, R. and {Vaduvescu}, O. and {St{\u a}nescu}, M.},
  title = {Distant trans-Neptunian object candidates from NASA's TESS mission scrutinized: fainter than predicted or false positives?},
  journal = {MNRAS},
  year = 2022,
  volume = {513},
  number = {1},
  pages = {L78--L82},
  doi = {10.1093/mnrasl/slac036},
  eprint = {2204.02230},
  archivePrefix = {arXiv}
}

@article{Nguyen2024,
  author = {{Nguyen}, Tam and {Woods}, Deborah F. and {Ruprecht}, Jessica and {Birge}, Jonathan},
  title = {Efficient Search and Detection of Faint Moving Objects in Image Data},
  journal = {AJ},
  year = 2024,
  volume = 167,
  pages = {113},
  doi = {10.3847/1538-3881/ad20e0}
}

@article{GeringerSameth2025,
  author = {{Geringer-Sameth}, Alex and {Golovich}, Nathan and {Iwabuchi}, Keita},
  title = {Multi-year stacking searches for solar system bodies},
  journal = {arXiv e-prints},
  year = 2025,
  pages = {arXiv:2509.25428},
  eprint = {2509.25428},
  archivePrefix = {arXiv},
  primaryClass = {astro-ph.EP},
  doi = {10.48550/arXiv.2509.25428}
}

@article{BernsteinKhushalani2000,
  author = {{Bernstein}, G. and {Khushalani}, B.},
  title = {Orbit Fitting and Uncertainties for Kuiper Belt Objects},
  journal = {AJ},
  year = 2000,
  volume = 120,
  pages = {3323},
  doi = {10.1086/316868}
}

@article{PravecHarris2007,
  author = {{Pravec}, Petr and {Harris}, Alan W.},
  title = {Binary asteroid population. 1. Angular momentum content},
  journal = {Icarus},
  year = 2007,
  volume = 190,
  pages = {250},
  doi = {10.1016/j.icarus.2007.02.023}
}

@article{ForemanMackey2013,
  author = {{Foreman-Mackey}, Daniel and {Hogg}, David W. and {Lang}, Dustin and {Goodman}, Jonathan},
  title = {emcee: The MCMC Hammer},
  journal = {PASP},
  year = 2013,
  volume = 125,
  pages = {306},
  doi = {10.1086/670067}
}

@article{Astropy2022,
  author = {{\unskip Astropy Collaboration} and {Price-Whelan}, Adrian M. and {Lim}, Pey Lian and others},
  title = {The Astropy Project: Sustaining and Growing a Community-oriented Open-source Project and the Latest Major Release (v5.0) of the Core Package},
  journal = {ApJ},
  year = 2022,
  volume = 935,
  pages = {167},
  doi = {10.3847/1538-4357/ac7c74}
}

@misc{Burke2020,
  author = {{Burke}, C.~J. and {Levine}, A. and {Fausnaugh}, M. and others},
  title = {TESS-Point: High precision TESS pointing tool},
  publisher = {Astrophysics Source Code Library, record ascl:2003.001},
  year = 2020,
  eprint = {2003.001},
  archivePrefix = {ascl}
}

@article{Woods2021,
  author  = {Woods, Deborah F. and Ruprecht, Jessica D. and Kotson, Michael C. and Main, Erin L. and Evans, Elizabeth W. and Varey, Jacob D. and Vaillancourt, John E. and Viggh, Herbert E. M. and Brown, Joshua J. and P{\'a}l, Andr{\'a}s},
  title   = {Asteroid Observations from the Transiting Exoplanet Survey Satellite: Detection Processing Pipeline and Results from Primary Mission Data},
  journal = {PASP},
  year    = {2021},
  volume  = {133},
  pages   = {014503},
  doi     = {10.1088/1538-3873/abc761}
}

@article{Holman2026,
  author  = {Holman, Matthew J. and Juri{\'c}, Mario and Napier, Kevin J. and Kurlander, Jacob A. and Heinze, Aren and others},
  title   = {2026 ED138: A Large Trans-Neptunian Object in Early Rubin Observatory Data, and How Many More May Remain},
  journal = {arXiv e-prints},
  year    = {2026},
  pages   = {arXiv:2610.38693},
  eprint  = {2610.38693},
  archivePrefix = {arXiv},
  primaryClass = {astro-ph.EP},
  doi     = {10.48550/arXiv.2610.38693}
}

@incollection{Muller2020,
  author    = {M{\"u}ller, Thomas and Lellouch, Emmanuel and Fornasier, Sonia},
  title     = {Trans-Neptunian objects and Centaurs at thermal wavelengths},
  booktitle = {The Trans-Neptunian Solar System},
  editor    = {Prialnik, Dina and Barucci, M. Antonietta and Young, Leslie},
  publisher = {Elsevier},
  year      = {2020},
  pages     = {153--181},
  doi       = {10.1016/B978-0-12-816490-7.00007-2}
}

@book{Bate1971,
  author    = {Bate, Roger R. and Mueller, Donald D. and White, Jerry E.},
  title     = {Fundamentals of Astrodynamics},
  publisher = {Dover},
  address   = {New York},
  year      = {1971}
}

@article{Bellm2019,
  author  = {Bellm, Eric C. and Kulkarni, Shrinivas R. and Graham, Matthew J. and others},
  title   = {The Zwicky Transient Facility: System Overview, Performance, and First Results},
  journal = {PASP},
  year    = {2019},
  volume  = {131},
  pages   = {018002},
  doi     = {10.1088/1538-3873/aaecbe}
}

@article{Chambers2016,
  author  = {Chambers, K. C. and Magnier, E. A. and Metcalfe, N. and others},
  title   = {The Pan-STARRS1 Surveys},
  journal = {arXiv e-prints},
  year    = {2016},
  pages   = {arXiv:1612.05560},
  eprint  = {1612.05560},
  archivePrefix = {arXiv},
  primaryClass = {astro-ph.IM},
  doi     = {10.48550/arXiv.1612.05560}
}

@article{Bannister2018,
  author  = {Bannister, Michele T. and Gladman, Brett J. and Kavelaars, J. J. and others},
  title   = {OSSOS. VII. 800+ Trans-Neptunian Objects---The Complete Data Release},
  journal = {ApJS},
  year    = {2018},
  volume  = {236},
  pages   = {18},
  doi     = {10.3847/1538-4365/aab77a}
}

@article{Petit2011,
  author  = {Petit, J.-M. and Kavelaars, J. J. and Gladman, B. J. and others},
  title   = {The Canada-France Ecliptic Plane Survey---Full Data Release: The Orbital Structure of the Kuiper Belt},
  journal = {AJ},
  year    = {2011},
  volume  = {142},
  pages   = {131},
  doi     = {10.1088/0004-6256/142/4/131}
}

@article{Bernardinelli2022,
  author  = {Bernardinelli, Pedro H. and Bernstein, Gary M. and Sako, Masao and others},
  title   = {A Search of the Full Six Years of the Dark Energy Survey for Outer Solar System Objects},
  journal = {ApJS},
  year    = {2022},
  volume  = {258},
  pages   = {41},
  doi     = {10.3847/1538-4365/ac3914}
}

@article{Masci2019,
  author  = {Masci, Frank J. and Laher, Russ R. and Rusholme, Ben and others},
  title   = {The Zwicky Transient Facility: Data Processing, Products, and Archive},
  journal = {PASP},
  year    = {2019},
  volume  = {131},
  pages   = {018003},
  doi     = {10.1088/1538-3873/aae8ac}
}

@article{ReinLiu2012,
  author  = {Rein, H. and Liu, S.-F.},
  title   = {REBOUND: an open-source multi-purpose N-body code for collisional dynamics},
  journal = {A\&A},
  year    = {2012},
  volume  = {537},
  pages   = {A128},
  doi     = {10.1051/0004-6361/201118085}
}

@article{Cerati2026,
  author  = {Cerati, Gioele},
  title   = {An Archival Search for Outer Solar System Objects in the NOIRLab Source Catalog DR2: 65 Candidates, Including Four L4 Neptune Trojans},
  journal = {arXiv e-prints},
  year    = {2026},
  pages   = {arXiv:2611.14330},
  eprint  = {2611.14330},
  archivePrefix = {arXiv},
  primaryClass = {astro-ph.EP},
  doi     = {10.48550/arXiv.2611.14330}
}

@article{Nidever2021,
  author  = {Nidever, David L. and Dey, Arjun and Fasbender, Katie and others},
  title   = {Second Data Release of the All-sky NOIRLab Source Catalog},
  journal = {AJ},
  year    = {2021},
  volume  = {161},
  pages   = {192},
  doi     = {10.3847/1538-3881/abd6e1}
}

@article{Flaugher2015,
  author  = {Flaugher, B. and Diehl, H. T. and Honscheid, K. and others},
  title   = {The Dark Energy Camera},
  journal = {AJ},
  year    = {2015},
  volume  = {150},
  pages   = {150},
  doi     = {10.1088/0004-6256/150/5/150}
}

@article{Dey2019,
  author  = {Dey, Arjun and Schlegel, David J. and Lang, Dustin and others},
  title   = {Overview of the DESI Legacy Imaging Surveys},
  journal = {AJ},
  year    = {2019},
  volume  = {157},
  pages   = {168},
  doi     = {10.3847/1538-3881/ab089d}
}

@article{Flewelling2020,
  author  = {Flewelling, H. A. and Magnier, E. A. and Chambers, K. C. and others},
  title   = {The Pan-STARRS1 Database and Data Products},
  journal = {ApJS},
  year    = {2020},
  volume  = {251},
  pages   = {7},
  doi     = {10.3847/1538-4365/abb82d}
}

@article{Magnier2020,
  author  = {Magnier, Eugene A. and Schlafly, Edward F. and Finkbeiner, Douglas P. and others},
  title   = {Pan-STARRS Photometric and Astrometric Calibration},
  journal = {ApJS},
  year    = {2020},
  volume  = {251},
  pages   = {6},
  doi     = {10.3847/1538-4365/abb82a}
}

@inproceedings{Boulade2003,
  author    = {Boulade, O. and Charlot, X. and Abbon, P. and others},
  title     = {MegaCam: the new Canada-France-Hawaii Telescope wide-field imaging camera},
  booktitle = {Instrument Design and Performance for Optical/Infrared Ground-based Telescopes},
  series    = {Proc. SPIE},
  year      = {2003},
  volume    = {4841},
  pages     = {72--81},
  doi       = {10.1117/12.459890}
}

@article{Gwyn2012,
  author  = {Gwyn, Stephen D. J. and Hill, Norman and Kavelaars, J. J.},
  title   = {SSOS: A Moving-Object Image Search Tool for Asteroid Precovery},
  journal = {PASP},
  year    = {2012},
  volume  = {124},
  pages   = {579},
  doi     = {10.1086/666462}
}

@article{Barbary2016,
  author  = {Barbary, Kyle},
  title   = {SEP: Source Extractor as a library},
  journal = {JOSS},
  year    = {2016},
  volume  = {1},
  number  = {6},
  pages   = {58},
  doi     = {10.21105/joss.00058}
}

@article{BertinArnouts1996,
  author  = {Bertin, E. and Arnouts, S.},
  title   = {SExtractor: Software for source extraction},
  journal = {A\&AS},
  year    = {1996},
  volume  = {117},
  pages   = {393},
  doi     = {10.1051/aas:1996164}
}

@article{Harris2020,
  author  = {Harris, Charles R. and Millman, K. Jarrod and van der Walt, St{\'e}fan J. and others},
  title   = {Array programming with NumPy},
  journal = {Natur},
  year    = {2020},
  volume  = {585},
  pages   = {357},
  doi     = {10.1038/s41586-020-2649-2}
}

@article{Virtanen2020,
  author  = {Virtanen, Pauli and Gommers, Ralf and Oliphant, Travis E. and others},
  title   = {SciPy 1.0: fundamental algorithms for scientific computing in Python},
  journal = {NatMe},
  year    = {2020},
  volume  = {17},
  pages   = {261},
  doi     = {10.1038/s41592-019-0686-2}
}

@article{Hunter2007,
  author  = {Hunter, John D.},
  title   = {Matplotlib: A 2D Graphics Environment},
  journal = {CSE},
  year    = {2007},
  volume  = {9},
  pages   = {90},
  doi     = {10.1109/MCSE.2007.55}
}

@inproceedings{Okuta2017,
  author    = {Okuta, Ryosuke and Unno, Yuya and Nishino, Daisuke and Hido, Shohei and Loomis, Crissman},
  title     = {CuPy: A NumPy-Compatible Library for NVIDIA GPU Calculations},
  booktitle = {Proceedings of Workshop on Machine Learning Systems (LearningSys) in The Thirty-first Annual Conference on Neural Information Processing Systems (NIPS)},
  year      = {2017},
  url       = {http://learningsys.org/nips17/assets/papers/paper_16.pdf}
}

@article{Holman2019,
  author  = {Holman, Matthew J. and Payne, Matthew J. and P{\'a}l, Andr{\'a}s},
  title   = {A TESS Search for Distant Solar System Planets: A Feasibility Study},
  journal = {RNAAS},
  year    = {2019},
  volume  = {3},
  pages   = {160},
  doi     = {10.3847/2515-5172/ab4ea6}
}

@article{Payne2019,
  author  = {Payne, Matthew J. and Holman, Matthew J. and P{\'a}l, Andr{\'a}s},
  title   = {A TESS Search for Distant Solar System Objects: Yield Estimates},
  journal = {RNAAS},
  year    = {2019},
  volume  = {3},
  pages   = {172},
  doi     = {10.3847/2515-5172/ab5641}
}

@article{Ganapathi2022,
  author  = {Ganapathi, Varun},
  title   = {Maximum Likelihood Systematic Effect Modeling and Matched Filtering to Detect Trans-Neptunian Objects with TESS},
  journal = {arXiv e-prints},
  year    = {2022},
  pages   = {arXiv:2209.09848},
  eprint  = {2209.09848},
  archivePrefix = {arXiv},
  doi     = {10.48550/arXiv.2209.09848}
}

@article{Alexandersen2026,
  author  = {Alexandersen, Mike and Holman, Matthew J. and Payne, Matthew J.},
  title   = {Cleaning TESS Images to Reveal Solar System Objects},
  journal = {PSJ},
  year    = {2026},
  volume  = {7},
  pages   = {192},
  doi     = {10.3847/PSJ/ae8a4d}
}

@article{Belskaya2003,
  author  = {Belskaya, Irina N. and Barucci, Antonella M. and Shkuratov, Yurij G.},
  title   = {Opposition Effect of Kuiper Belt Objects: Preliminary Estimations},
  journal = {EM\&P},
  year    = {2003},
  volume  = {92},
  pages   = {201},
  doi     = {10.1023/B:MOON.0000031938.00224.32}
}
\bibliographystyle{aasjournalv7}

\end{document}